# Events and the Ontology of Quantum Mechanics

Mauro Dorato
Department of Philosophy, Communication and Media Studies
University of Rome 3
Via Ostiense 234, 00146, Rome

Abstract

In the first part of the paper I argue that an ontology of events is precise, flexible and general enough so as to cover the three main alternative formulations of quantum mechanics as well as theories advocating an antirealistic view of the wave function. Since these formulations advocate a primitive ontology of entities living in four-dimensional spacetime, they are good candidates to connect that quantum image with the manifest image of the world. However, to the extent that some form of realism about the wave function is also necessary, one needs to endorse also the idea that the wave function refers to some kind of power. In the second part, I discuss some difficulties raised by the recent proposal that in Bohmian mechanics this power is holistically possessed by all the particles in the universe.

Key words: events; primitive ontologies, holistic powers

1. Introduction

The main thesis of this paper is that events are the most suitable metaphysical category that we should use to describe the metaphysical intimations of quantum mechanics.[1] The paper is divided into two parts, which also correspond to the two main arguments that I will provide in support of my thesis.

More in details, in section 2 of the first part I show the centrality of the metaphysical category of events in the ontology presupposed by ordinary language and commons sense. Section 3 argues that the this metaphysical category is precise, flexible and general enough so as to cover the three main alternative to standard quantum mechanics that advocate a primitive ontology of entities living in four-dimensional spacetime – the flash and the density-of-matter versions of GRW and Bohmian mechanics –, as well as some antirealistic views about the wave function that are presented in section 4. Consequently, an ontology of events is the best candidate to bridge the gap between the manifest and the physical image of the world emerging from quantum mechanics. Given the many non-intuitive features displayed by the latter, this result is philosophically important.

---

[1] See Auyang 1995, Dieks 2002 and Haag 2012 for a similar conclusion based on quantum field theory.

The remaining question is then whether events *suffice* for a clarification of the ontological assumptions of the views of quantum mechanics discussed in the first part, or some form of wave function realism is also needed. As an attempt to answer this question, the second part of the paper (section 5) focuses on quantum dispositions: considering their role in the ontology presupposed by ordinary language, which is full of dispositional terms, also dispositions are an obvious candidate for the wished-for rapprochement of the two images mentioned above. However, while dispositions or powers seem a plausible basis for defending some kind of nomological realism about the wave function, Esfeld et al. (2013) have recent argued that the wave function ought to be regarded as describing a power of the configuration of all particles in the universe. This claim raises the question whether the primitive ontology of quantum theory can plausibly be given in terms of matter points devoid of any intrinsic dispositions but essentially identified by metrical relations (Esfeld, et al. 2015).

2. Events as the key bridge between the physical and the manifest image of the world

One of the most frequent points of misunderstanding between physicists and philosophers of physics or metaphysicians is not only caused by differences in language but also by the fact that philosophers worry much more than physicists about *ontological* issues, namely interpretive questions involving what (typically a poorly understood) physical theory tells us about the world. In the case of quantum mechanics, however, interpretive questions calling for ontological analyses (" "how could the world be like if quantum theory is true?") become murky since, at least according to philosophers, it is still controversial how quantum theory should be formulated, given that there are various proposals in the market

A more wide-ranging and fruitful way to pursue the philosophy of physics has been beautifully expressed by Suppes, and as an tribute to his intellectual legacy, I quote a passage from his self-profile: "It is my conviction that an important function of contemporary philosophy is to understand and to formulate as a coherent world view the highly schematic character of modern science and the highly tentative character of the knowledge that is its aim." (Suppes 1989, p. 47). Examining the relation between the physical and the manifest image of the world in Sellar's sense (1962) is a rather promising way to try to achieve Suppes' aim of coherence, and it is the path on which I want to venture here.

In my opinion, in order to achieve Suppes and Sellar's aim, in our case one needs to provide:

1. A precise ontology for a precisely specified quantum theory *T*;
2. A clear description of the aspects of the manifest image that are relevant for *T*. Given the vagueness of the "manifest image", philosophers need help from the cognitive sciences, from conceptual analysis and descriptive metaphysics (Strawson 1959) and from whatever else is deemed to be related to the task at hand. Finally, and mostly importantly, one must:
3. Put 1 and 2 together, showing how 1 relates to 2. In case of conflict, some explanation is needed to remove it in order to create a greater epistemic unity and harmony in our worldview.

It seems to me that it is mainly the presence of 2 and 3 that justifies philosophers in embarking in the perilous waters of ontology suggested by 1, so as "to understand *how things* in the broadest possible sense of the term *hang together*, in the broadest possible sense of the word" (Sellars 1962, p.1).[2]

As a general remark, and in order to begin our discussion, it can hardly be denied that physics has replaced most of our ordinary-language concepts by transforming them into precisely defined notions: work, heat, energy, light space and time are only some examples. And, of course, it has invented wholly new concepts, which have helped to overcome some false presuppositions of commonsense implied by our cognitive makeup or by our reliance on ordinary language.

However, the strategies of "precisification" and "replacement" of common-sense concepts have their important exceptions and the notion of "*event*" is possibly the most important one. By belonging essentially to ordinary language *and* to spacetime theories, *events* – even if idealized, as they are in the latter, where they are treated as *pointlike* – are excellent candidates to connect Sellars' two images. The main reason, I submit, is that *events* (and their happening in temporal succession, originating *processes*) play a major role in making sense of the passage of time. All "historical" sciences (cosmology, astrophysics, geology, evolutionary biology, individual psychology and all the social sciences) essentially refer to *events that happen in temporal succession*, or processes. Such *de facto* irreversible sequences among time-like separated events is the main reason to believe in an objective passage of time, an essential feature of the natural and the social world, as well as of our inner mental life. If spacetime physics *presupposes* some concepts coming from ordinary language, then the *meaning* of such concept must be taken into account. Now, "events" *occur* by definition, their "being" is their happening, and they *are* an essential ontological component of any spacetime theory; if becoming *is* – albeit *very* minimally and controversially – given by the successive, irreversible, worldline-dependent occurrence of events (Savitt 2001, Dieks 2006, Dorato 2006,), then there is objective, local becoming also in the special and in the general theory of relativity.

---

[2] Trying to establish coherence between science and intuitive assumptions about the world forced upon us by our hardwired cognitive makeup points to the ideal of unified science recently defended by Ladyman and Ross (2007).

Given that events originating processes are an important ontic category of both spacetime physics and the manifest image (in the minimal sense of "happenings in a temporal order" specified above), a natural question at this point is: what is the role of events in the ontology of quantum mechanics, given the non-separability of measurement outcomes and the failure of supervenience of relations of entanglement on relata?

In order to answer these questions, Meyer's schematic classification of theories of events is rather useful (Meyer 2013, p.14).[3] The first theory identifies events with *regions of spacetime* (Quine 1985 and Lemmon 1967). The second reduces them to *properties* of spatiotemporal regions (Lewis 1986). The third identifies events with properties $P$ instantiated by substances $S$ (physical systems) at some time (Kim 1976); since substances are located in space, Kim's definition ought to be supplemented by adding a location in space to instants of time, so as to render his account compatible with relativity. The fourth view identifies events with states of affairs instantiated in a spatiotemporal region by an object $S$ (Chisholm 1971).

Here I will not try to adjudicate among these views – they are not exhaustive – or to determine whether the metaphysical category of *objects* and their properties is more fundamental than that of events, as all of them seem to imply.[4] In what follows I will simply take for granted that primitivism about events is not an incoherent view, so that (i) objects $S$ (physical systems) *can be* regarded as more or less macroscopic and as more or less monotonous collections of events, and (ii) dispositional properties can also be grounded in events, since a disposition needs at least in principle not only a stimulus (the striking of the match) but also the possibility of a manifestation (its burning), which are both events.

For my purposes it is essential to note that *all* of these theories of events refer or presuppose, *spacetime*.[5] Therefore, events seem very promising candidates to clarify the ontological implications of theories of quantum mechanics that assume that the latter is primarily about spatiotemporally extended "stuff" (Allori et al. 2008, Allori 2013). If this claim turned out to be validated by the next sections, we could conclude that events play a crucial role in establishing the hoped-for connection between the world of our experience and our fundamental theory of the physical universe.

---

[3] Here, I will simplify it to adapt it to my purpose.
[4] Primitivists about properties claim that objects are collection of properties, while events are simply exemplifications of properties, while primitivists about objects regard them as the ground for properties and events. Pluralistic views are against these forms of monism.
[5] Also van Bentham's view (1983), which claims that events are *times* during which certain statements hold, need time (and space, I add, given relativistic injunctions) as essential ingredients.

3 Events in a primitive ontology of spatiotemporally extended entities

The main theories of quantum mechanics advocating a primitive ontology of spatiotemporally extended stuff are Bohmian mechanics and the two versions of GRW dynamical collapse models – namely the so-called "flashes" version (Tumulka 2006) and the field version, based on processes of mass density localizations (Ghirardi Grasso and Benatti 1995).

These theories are in contrast with those interpretations of quantum mechanics that take its formalism at face value. Among these, we find the view that the ontology of quantum theory is given either by the *quantum state of the universe* (see Saunders et al 2010 for a sample of papers defending this view) or by *configuration space* (Albert 1996, Ney 2013). Theories advocating a primitive ontology of spatiotemporal entities (Allori et al 2008, Maudlin 2010), on the contrary, consider the wave function, and the corresponding quantum state denoted by it, at most as an element that determines the dynamical evolution of the primitive ontology. For this reason, by using a potentially misleading expression that, however, stuck, the wave function is often dubbed "a nomological entity" (Dürr, Goldstein, and Zanghì 2013).[6]

With respect to the flash version of GRW, Bell himself argued that its ontology is *described* by jumps of the wave functions in configurations space that, in the real world, correspond to events localized at points of spacetime, the local beables of the theory: "The GRW jumps… are the mathematical counterpart in the theory to real events at definite places and times…a piece of matter is a galaxy of such events" (Bell 1989, p. 205). In the flash ontology, the physical world is discrete, in the sense that, unlike the case of Bohmian mechanics to be discussed below, there are no continuous worldlines traced by particles, but only events scattered in spacetime; for this reason, spacetime is mostly empty, in the same sense in which, according to physics, a chair is almost empty since it is made of atoms whose mass is mainly concentrated in the nuclei. In order to recover the existence of ordinary objects of the manifest image from the discrete ontology of the quantum image consists in supposing that there are many more flashes wherever those objects are located than there are in other regions. As to the dynamical law coded by the wave function, flashes/events at given spacetime points, together with the wave function, determine the probability for the next flashes/events to occur at other place-times. It is not implausible to attribute these events the intrinsic probabilistic disposition to determine the occurrence of other flashes (Dorato and Esfeld 2010), so that the role of the law can be absorbed by these dispositions *à la* Mumford (2004). It will be important later on to discuss the possibility of extending these intrinsic dispositions also to particles in Bohmian mechanics (see Brown et al. 1996, Esfeld et al. 2015). In a word, there is little doubt that an ontology of events is very much at home in the "flashy" version of GRW.

---

[6] Here I should note that I do not defend any of these theories, but I am only exploring their metaphysical commitments.

In the mass-density version of GRW, the situation looks at first rather different. In this version in fact, the collapses of the wave function in configuration space do not represent an "atomistic" world of localized events subject to chancy Lucretian *clinamena* ("swerves"),[7] but rather the localizations of a continuous field constituted by the *density of matter* in physical space. In other words, in this case the spontaneous hit of the wave function refers to an increase of the "density of stuff" in a spacetime region. Ordinary objects are redescribed in terms of regions of space in which there is a significant increase in the density of stuff. Given the *discrete* nature of events, however, how can we square their apparent atomistic ontology with the continuous, field-like gunk postulated by the density-of-matter version of GRW?

In order to answer this question, one should first of all notice that the *contraction* of the matter field – as described by the localization of the wave function in configuration space – is certainly an event, as it is a *change* in the density, and all changes are events. Furthermore, this contraction is: i) either all there is or ii) a particular event in an evolving set of events or process or iii) a change in a preexisting substance. Option i) needs no discussion, since in this case the primitive ontology of the theory would be given just by events of "contractions", in the same sense in which, in the flash version of GRW, flashes/events are all there is.

The difference between ii) or iii) depends on how we want to characterize the continuous matter field in which the events of contraction occur. According to the standard definition of a *classical* field, a field is a continuous distribution of physical quantities (scalar, vector fields or tensor fields as they may be) over spacetime points or spatiotemporal regions. An electromagnetic field in a region of spacetime, for instance, is a continuous assignment of the relevant magnitudes of the electric and magnetic fields at the spacetime points of the region in question; these points are a set of events describable or identifiable by certain well-defined magnitudes of the field.

Going now to the discussion of option ii), each of the points in the region *is* an event, and the region is a set of events that instantiate different properties. Also in the fourdimensional version of option (ii), spatiotemporal regions can be described as covered by an infinite number of worldlines, and therefore by a succession of events or processes instantiating different properties, in the same sense in which a party (an event) is boring in its first part and exciting in its second part. As to option iii), which regards a field as a substance instantiating properties at spacetime regions, we can either adopt Kim's definition and identify as above a field with a set of events, or re-conceptualize substances by considering them to be big collections of events (see note 3). Also in this case, the desired conclusion that matter fields can be regarded as constituted by sets of events holds.

---

[7] The first to compare GRW's collapses to Lucretius' indeterministic philosophy was van Fraassen's (see Ghirardi 2005, p. 424).

There is no reason why this result, obtainable for classical fields, should not apply to GRW's *quantum* fields of matter density. Summarizing what was stated above, a field can be specified by a time-dependent mapping from each point of space to a field value φ(**x**,*t*). While in the case of relativistic quantum fields the mapping is between a spacetime region and the value of an *operator* – so that the field is question is not a specific assignment of physical magnitudes to spacetime points or regions – a non-relativistic quantum field in GRW's sense *can* be regarded as an assignment of a definite value of density of matter to a region or to the entirety of spacetime, a constraint that seems necessary to have a "kosher" field (see Kuhlmann 2014). According to the minimalistic view of relativistic becoming specified at the bottom of p.3, this field, is a set of events evolving in time, and suffers contractions and localizations with the passage of (absolute) time depending on the dynamics specified by the wave function.

Finally, consider the various theories of events mentioned above: all of them accommodate quite well the two theories mentioned above. According to Quine and Lemmon's view of events, we can claim that a spacetime region – in our case occupied by the density of matter field – just *is* a continuously distributed collection of physical events. Notice furthermore that Lewis' approach to events as regions of spacetime could fit easily as well for the case of a GRW field: the properties of a spacetime region (i.e., events) are the field itself. Likewise for Kim's exemplification doctrine, in which events are properties exemplified by physical systems S at spacetime points (s, t): the exemplification of the density of stuff (property) at points (s,t) by a physical system S (the field) is an event. It seems that we can conclude that the difference between a discrete and a continuous ontology in the case of the two versions of GRW is not sufficient to exclude the attribution of an ontology of events also to the continuous field version.

Finally, we must now extend these results to Bohmian mechanics' s primitive ontology, given by continuous worldlines traced by particles in spacetime. If particles at points are identified with physical, pointlike events, worldlines can be regarded as a continuous collection of such events in space-time that represents the history of the particles: these worldlines are therefore *processes*. It then seems that an ontology of temporally ordered events or processes can also ground the continuous collection of worldlines traced by Bohmian particles, which constitute the primitive ontology of the theory.

3.1 Events as materia prima?

In a forthcoming publication, (Esfeld et al. 2015) identify the primitive ontology of Bohmian mechanics (and of the other two versions of GRW) with *substances* (matters points) devoid of intrinsic properties that are essentially identified by metrical *relations*.[8] It is important to compare Esfeld's neo-Cartesian, structuralist view of quantum mechanics (Esfeld et al. 2015) with the view that is defended here. First of all, matter points can be easily replace with pointlike, physical events (the ontology that is proposed here). If in addition I claimed, as they do, that (i) spatiotemporal relations are constructed, relationally, as identity-providing metrical relations holding between *events*; if (ii) events were identified, as I proposed in discussing the density of stuff version of GRW, with substances, and if (iii) events were devoid of any intrinsic power or property, then the difference between the ontology of events proposed here and that defended by Esfeld et al. (2015) would be purely *verbal*.

However, despite some very important similarities stemming from the adoption of a privileged spatiotemporal perspective, this conclusion is not be completely correct. By concentrating on (iii), I will now show that there are at least two points that could mark the difference between the ontological view defended here and that advocated in the above paper.

1. An ontology of events has a more general applicability than the rather sparse structuralist view of quantum mechanics advocated by Esfeld and co. Events in fact are the essential ontological ingredient also in antirealistic views of the wave function, which cannot be ruled out by *fiat*, and which need no structuralist interpretation of the primitive ontology. Note that this is *not* a criticism of Esfeld et al. (2013, 2015), since they do *not* consider antirealistic views of quantum mechanics. Of course, the fact that an ontology of events is applicable also to the latter views (views that here I am not endorsing) would not count as an advantage if wave function antirealism could be ruled out. Unfortunately, since this is not the case, an event ontology has an additional advantage with respect to materia prima since, as we will see in the next section, it helps to illustrate in a very perspicuous way a persisting risk of GRW and Bohm. The risk – which should not discourage further pursuits of these formulations – is given by the fact, to be discussed in the next section, that their greater explanatory power so far lacks an independent confirmation given by new predictions.

Of course, if the standard view of quantum mechanics were *inconsistent* the event ontology would not be really an "advantage". In this case, the standard view would not be acceptable as a scientific theory because it would not explain the definiteness of the properties of the macro-objects of the manifest image (if quantum theory is complete and the evolution of the quantum state is always linear). But this criticism is correct only if the wave function can be given a realistic interpretation of some kind, an interpretation that is not without difficulties and that has been denied by views of Bohmian mechanics that regard $\Psi$ as a nomological entity.

---

[8] This is a form of moderate structural realism, since relata exist even while being devoid of an intrinsic essence. As such that are individuated via a structure.

2. The second point raises some questions to readings of Bohmian mechanics that enjoins us to "pack" under the universal wave function *all* putatively intrinsic properties of elementary particles, like mass, spin or charge. This reading goes against a "generosity" principle advocated by Brown et al. (1996), Pylkkänen et al. (2014), who plausibly attribute particles some degree of intrinsicality. If the wave function really "embodied" all apparently intrinsic properties of particles, and referred to a global power of the configuration of particles, how could such a global power be considered concrete and not just an abstract property of the configuration? The fact that the wave function is a nomological object seems to exclude the possibility that mass and charge reside completely in it, because they would become abstract as well.

The first point will be illustrated in the next section (4), while the second in the final section 5.

4 Events in some antirealistic views of the wave function

One of the crucial questions raised by contemporary philosophers of quantum mechanics involves the ontological status of the wave function. One way to formulate the question is as follows:

Q) *Is the austere ontology of events presented above sufficient to make sense of quantum mechanics?*

Roughly speaking, according to wave-function realists, the answer is in the negative: the wave function denotes some real feature of the physical world: a primitive quantum state (Maudlin 2013), a nomological "*entity*" (Goldstein and Zanghì 2013), some property instantiated by all particles in the universe (Monton 2006, Esfeld et al 2013), a physical 3N-dimensional configuration space (Albert 1996, Albert and Ney 2013) or all there is, as in the many-world interpretations (Wallace 2014). In a word, according to this negative answer, events – while being necessary components of the primitive ontology – are however not sufficient to give a complete account of the ontological intimations of quantum mechanics.

According to antirealists, as we are about to see, the answer to the question above is positive. For the sake of brevity, here I will mention only two main brands of wave function antirealism, and two representatives for each "brand". In one of these antirealistic brands, measurement outcomes (Bohr 1972-2006), or interactions between quantum systems (Rovelli 1996), constitute the main ontology of quantum theory and therefore of the quantum world. In the second brand, an ontology of events is suggested by difficulties generated by a particle ontology (Haag 2013, Pashby 2014).

Even a brief sketch of these four positions will suffice to drive home my main point.

According to Bohr, the ontology of the quantum world contains mind-independent elementary particles (as in entity realism)[9], but by excluding any realistic status to the wave function and therefore to collapses, it admits only *macroscopic* measurements outcomes, which are the results of non-controllable, nonseparable interactions between microscopical particles and macroscopical, classical measurements apparata. The outcomes are *events* with a precise location in space and time: the microscopic or macroscopic nature of these events depends contextually on the kind of measurement one performs.

Rovelli's relational quantum mechanics (RQM), on the other hand, contaminates Bohr's view with Everettian quantum mechanics, from which it is distinguished by a significant quantum-state antirealism. According to RQM, the quantum ontology is constituted by events, regarded as outcomes of interactions between *any* two systems, macroscopic or *microscopic* as they may be. In Rovelli's RQM, the outcomes of any interaction (micro-micro or micro-macro) is a definite event. "the real events of the world are the 'realizations' (the 'coming to reality', the 'actualization') of the values . . . in the course of the interaction between physical systems. These quantum events have an intrinsically discrete (quantized) granular structure." (Rovelli 2005, p. 117). Consequently, an interaction between a Stern-Gerlach apparatus measuring spin up and a particle is in fact the outcome of the magnification of a microscopic but property-definite *event* that is the outcome of an interaction between the quantum particle and the microscopic component of the Stern-Gerlach apparatus that interacts with it. In this sense, microscopic or macroscopic events are the primitive ontology of RQM (see Rovelli 1996, Laudisa and Rovelli 2013, Dorato 2015). According to RQM, as in Bohr, attributing definite states to non-interactive physical systems has no meaning.

To go back to the question in italics raised at the beginning of this section, in these two accounts – among many others, and despite their significant differences – the wave function represents only the statistics of previous measurement outcomes, and is basically a bookkeeping, merely predictive device.

Now if we compare the ontological assumptions of the two versions of GRW with Rovelli's RQM we realize that *both are based on events as primitive ontology*. But if we add to the ontic interpretation defended by RQM the precise, explanatory theory of the "interaction" between two systems offered by the former, we get either GRW's flashes or the contractions of the density of the matter field! Interestingly, the difference between relational quantum mechanics and GRW flashes is crucial to understand the philosophical and methodological dissimilarity between Rovelli's (or, in a different sense, Bohr's) interpretation and the two GRW models. The latter, unlike the former, makes some precise assumptions about "where", "how", when and "why" a collapse occurs. This provides a non-contextual solution to the measurement problem and gives a clear theoretical description of the notion of "interaction" (Rovelli) or "measurement" (Bohr), which these two physicists regard as primitive or indescribable.

---

[9] Bohr can be considered a entity realist, since he believes in the mind-independent existence of electrons, protons, atoms and so on. However, he is an antirealist about quantum theory because he denies the reality of the wave function (Faye 1991). The same split realism can be attributed to Rovelli (Dorato 2014).

Unfortunately, so far at least, there is no *independent empirical evidence* that can be added to the remarkable explanatory power of the dynamical reduction models (and of Bohm's theory): GRW and Bohm lack some factual evidence that is independent of the fact that, *if* they were true, they would solve the measurement problem. Even though dynamical reduction models are, unlike Bohmian mechanics, testable in principle (Ghirardi and Bassi 2003), the fact that they have not been tested yet boosts the working physicists' skepticism and their subsequent, exclusive focus on the *final outcomes* of the reduction process, namely events, whose postulation is common to the two theories. The case of the mechanical models of the ether pursued in the 19$^{th}$ century (Darrigol 2000, Shaffner 2004) should raise some skepticism about theories that have a remarkable explanatory power but no independent evidence in their favor. In our case, since there is an indefinite numbers of different explanations of why we get definite results out of entangled particles interacting with measurements apparata (Bohm's, GRW's, decoherence processes in many-worlds'), we might (for the time being at least) follow RQM and regard the interaction described by the Hamiltonian of the two system as primitive and therefore in need of no explanation: after all, scientific revolutions are characterized by the fact that what needs to be explained before and after the revolution changes (Kuhn 1970, p. 104, and Kuhn 1977, p. 29)..

A second brand of antirealism about the wave function depends on the claim that any talk of "properties possessed by physical systems" (their position in particular) in quantum theory is inadequate, and must be replaced by an ontology of events localized in certain region of spacetime. Despite the different motivations, the end result is not too distant from the two positions we have discussed so far: "What do we detect? The presence of a particle? Or the occurrence of a microscopic event? We must decide for the latter… The detector fulfills two functions. It offers a target for a collision process, a microevent which is almost always the ionization of some molecule in the detector. Secondly it gives the amplification to visible dimensions via a chain reaction." (Haag 2013, p.1310). The main reason for this conclusion is that an event can have a precise location in space and time if and only if the momentum and energy transfer during the amplification mechanism are remarkably great: "… the operational approach to a point (in fact a point in space-time) can only be realized by a high-energy interaction process involving several particles—an event. The notion of "events" must be regarded as an independent primary concept intimately tied to relations in space-time (Haag 2013, p. 1298).

In a word, though for very different reasons, analogously to the three primitive-ontology theories discussed in the previous section, also according to Haag events are the main ontological assumption of quantum theory. In Haag as in RQM however, the label "event" applies not just to macroscopic, but also to microscopic processes, as those that occur in ionization.

In Pashby's account (2014), the problem leading to an ontology of events is given by the fact that a particle cannot be localized in a spatial region at a time. Consider an interval of time $\Delta(t_0, t_1)$ during which a particle $S$ is localized in a spatial region $R$: as a consequence of a theorem by Hegerfeldt (1998) that he mentions, it follows that if $S$ is in $R$ at *each* instant $t \in \Delta$, then $S$ is in $R$ at all times (Pashby 2014). Now take the contrapositive of this conditional. Suppose that there is a time at which $S$ is *not* located in $R$, i.e., suppose that outside the short present temporal interval $\Delta$ there is a future time $t_2$ at which $S$ is located in a region $R' \neq R$ (say, as in Pashby's example, that $R'$ is the region outside the lab $R$). Then $S$ cannot be located in the lab now (i.e. in $\Delta$) except in the case in which the measure of $\Delta$ is zero: "But what sort of persisting physical object fails to have spatial properties (in the regions we care about) at the vast majority of times? … this interpretation of the state, as describing the changing properties of a physical thing, is a mistake. In its place, I propose an account of localization in terms of the occurrence of spatio-temporally located events rather than possessed properties".

Summarizing the main point of this section, the preferability of an ontology of events over the minimalistic one proposed by Esfeld et al. (2015) depends on two facts. 1) Events are *necessary* both in realistic and in antirealistic views of the wave function: in virtue of their interpretation-independence, events turn out to be a central ontological component of quantum mechanics; at the same time, an ontology of bare matter points identified by metrical relation seems too sparse to be worthy of that name; 2) in virtue of its independence of particular theories of quantum mechanics, an ontology of events is capable to shed important light on the methodological costs that have to be paid to go beyond the antirealistic views, which are given by the fact that so far we have no uncontroversial *empirical* reason to endorse one among the various alternative theories of quantum mechanics presented in the previous sections.

Among the views that accord the wave function more than a predictive role, in the remainder of this paper I will concentrate on dispositionalism, which is the philosophical positions that regards the wave function as referring to a law-grounding power possessed by all entities living in the quantum world (Esfeld et. 2013).

## 5 Events and holistic dispositionalism: some concluding questions

I will now argue that if the primitive ontology, as advocated in Esfeld et al. (2015) is mere *res extensa* individuated structurally by metrical relations, and if the wave function in Bohmian mechanics refers to a global power of the whole configuration of particles of the universe (Esfeld et al. 2013, Esfeld et al. 2015), it is unclear how this power relates to the individual local substances, if these are devoid of any intrinsic power. Since the criticism in this section does not intend to be a definite rejection of this interesting proposal, it is meant to raise some questions that I think need to be answered in order to lend it more credibility.

The existence of a global power of the configuration of particles is a primitive fact in this proposal, so that questions about its origin would be misplaced. Therefore, if one questioned its intelligibility by asking how can it arise if the individual, pointlike matter points/particles are reduced to "substances" (?) with no *intrinsic power*, one would be left to wonder. Someone who found this power indigestible, however, could grant each particle *P* the *intrinsic* power to originate a holistic, irreducible power to fix each particle velocity in virtue of the position in the metric field of all *P*s. The irreducibility in question of course depends on the fact that in Bohmian mechanics the position of one particle alone would not suffice to fix the velocity of another one.

To illustrate this alternative account, consider the case of Bohmian spin. Esfeld et al. (2015) write "For a Bohmian particle to have "spin up" or "spin down" means nothing more and nothing less than … to *move* – in the pertinent measurement-context – in the respective way." As is well known, "spin" is a contextual relation holding between a Bohmian particle's position and the orientation of the apparatus. So if one were to assign a disposition to the singlet state and the Bohmian particle to display a certain measurement result, one should regard the disposition in question as *reducible* to relational properties between definite positions of particles and detectors. And yet, we can talk about an intrinsic, even though reducible, disposition of the particle to show a certain spin in one direction in certain measurement contexts (Clifton and Pagonis 1995, Dorato 2007).

The point I want to urge is that it is possible to claim that the individual particle has no definite spin in a given direction, but that it manifests an *intrinsic* disposition or has a concrete power to display spin up or spin down by interacting with the Stern-Gerlach apparatus (the stimulus of the disposition). In order to defend a perfect analogy between a Bohmian disposition "to have spin in a given direction" and classical but *intrinsic* dispositions like the fragility possessed by a glass, one could simply point out that the latter's "intrinsic disposition to break in a certain context" corresponds to "the intrinsic disposition to manifest a definite spin in a certain measurement context". Anyone committed to the reality of dispositions would clearly maintain that such a disposition is possessed by Bohmian particles also before measurement, and independently of it. Consequently, in the Bohmian case at least, the correct analogy with the classical disposition "fragility" is of course not given by "having a definite spin" but by "the disposition to have a definite spin in a given direction".

Without assigning localized matter points or particles *some* kind of *intrinsic* dispositions (as in Brown et al. 1996), to contribute (*together* with the other particles' disposition) to the emergence of an irreducible holistically power to fix the velocity of any particle in the global configuration, the version of holistic primitive ontology that Esfeld et al. (2015) paper defend might become unintelligible. It is for this reason that an ontology of events endowed with intrinsic dispositions but related by irreducible and holistic relations of entanglement seem preferable to an ontology constituted by primitive matter points devoid of any intrinsic powers but kept together so to speak by a holistic disposition.

A final difficult with Esfeld's reading is this: if matter points had no intrinsic dispositions (which does *not* entail having definite physical magnitudes), a subsystem described by an *effective* wave function would lack any intrinsic power whatsoever. But this would leave totally unexplained the dynamical behavior of the matter points located in our lab. This difficulty should incline the friends of dispositionalism, who believe in the fact that laws are made true by powers, to assign all particles intrinsic dispositions (not definite mass, but the disposition to have that mass in the context of manifestation relative to a subsystem). Agreed: the effective wave function is an arbitrary, pragmatically motivated *cut* of the whole universe in one of its subsystems and the rest of the universe. And the non-locality of the theory forces us to consider the rest of the world as contributing to the results due to the effective wave function. As Goldstein and Zanghì put it rather clearly, physicists always deal with subsystems because the wave function of the universe is epistemically inaccessible, but since the only fundamental object in Bohmian mechanics is the quantum universe, in order to draw any metaphysical conclusion it is natural to look only at the quantum universe.

And yet, consider the wave function $\psi(x)$ of a subsystem of the universe, where $\psi(x) = \Psi(x, Y)$ and Y the environment. The evolution of the positions of the particles *in* the subsystem must depend on laws too, and therefore on dispositions either intrinsically possessed by all particles in the subsystem, or holistically possessed by the whole configuration of the universe as in Esfeld's analysis. However, Goldstein and Zanghì (2013, p. 95) insist on the fact that if the subsystem is effectively *decoupled* from the rest of the environment Y, then the evolution of the subsystem follows just the Schrödinger' equation and the evolution can simulate or just display collapse. My point is simply that in this particular case, namely when physicists deal with subsystems as it is commonly the case, in order to make sense of the masses of the particles in the Hamiltonian of the Schrödinger equation it is necessary to attribute an intrinsic dispositions to all the masses of the particles of the subsystem:

$$H = - \sum_{K=1}^{N} \frac{h^2}{4\pi^2 2m_k} \nabla^2 + V$$

despite the nonlocal behavior of the wave function when it collapses. In a word, the problem with Esfeld and co.'s proposal is that it needs to alter the ontology when it moves from the configuration of the universe to the effective wave function: in my proposal, by attributing intrinsic dispositions to events in both cases, one need not change one's metaphysical assumptions about the nature of quantum dispositions.

Be that as it may, I want to conclude that given the important role played by events and powers in the ontology of quantum mechanics it is important to do future work on understanding their relationship in more details. A question that remains to be investigated is in fact whether are they independent of each other or is it possible to advance a more parsimonious metaphysical theory that unifies both.